\documentclass[letterpaper]{jpconf}

\usepackage{epsfig}
\usepackage[latin1]{inputenc}
\usepackage{amsmath}
\usepackage{amsfonts}
\usepackage{amssymb}
\usepackage[english]{babel}
\usepackage{slashed}

\newcommand{\nn}{\nonumber}
\newcommand{\be}{\begin{equation}}
\newcommand{\ee}{\end{equation}}
\newcommand{\bea}{\begin{eqnarray}}
\newcommand{\eea}{\end{eqnarray}}
\newcommand{\lp}{\left(}
\newcommand{\rp}{\right)}

\begin{document}

\title{Effective lagrangians for light-light interaction with a background field}

\author{J Mond\'ejar}

\address{Department of Physics, University of Alberta, Edmonton, Alberta, Canada T6G 2G7}

\ead{jmonde@phys.ualberta.ca}

\begin{abstract}

We address the issue of light-light scattering in the presence of a background field at low energies using effective lagrangians. We derive the Euler-Heisenberg lagrangian at one loop and modify it to incorporate the case of the interaction with a background field. 

\end{abstract}

\section{Introduction}

It is known since 1935 that vacuum polarization induces non-linear corrections to Maxwell's equations. Euler and Heisenberg \cite{EH} found an expression for the effective lagrangian describing light-light scattering for slowly varying fields at one loop, whose weak-field expansion at leading order is known as the Euler-Heisenberg lagrangian (although it was presented first in a former paper by Euler and Kockel \cite{Kockel}). Schwinger obtained years later an alternative expression for the effective lagrangian \cite{Schwinger}, and more recently the two-loop result was obtained \cite{Ritus1,Dittrich1,Fliegner,Dittrich2,Ritus2}. We have computed the Euler-Heisenberg lagrangian at one loop and modified it to incorporate the case of the interaction with a background field.

Recently, Penin \cite{Penin} studied the effects of Quantum Electrodynamics (QED) corrections on the Quantum Hall effect, and Dominguez et al. \cite{Dominguez} considered the field created by a charged spherical shell in the presence of an external, constant magnetic field. Although both papers deal with background field-related QED corrections, the methods employed are quite different. Our effective lagrangians provide some contact between these two papers. We start by presenting the Euler-Heisenberg lagrangian in section \ref{EH1}, and in section \ref{B} the lagrangians for the interaction with a background field. In section  \ref{P} we write down the expression for the polarization tensor in the presence of an external field, which connects with the calculations of Penin, and finally in section \ref{D} we redo the calculations of Dominguez et al. We present our conclusions in section \ref{conc}.

\section{The Euler-Heisenberg lagrangian}
\label{EH1}

The Euler-Heisenberg lagrangian is the first correction to the Maxwell lagrangian for weak fields. In this context, weak means $E, B \ll m^2/e$, where $m$ is the mass of the electron. That is, the Euler-Heisenberg lagrangian is a low-energy effective lagrangian, given by the first order in an expansion in the momentum of the external fields of the following diagram
\begin{displaymath}
\parbox{30mm}{
\includegraphics[width=0.29\columnwidth]{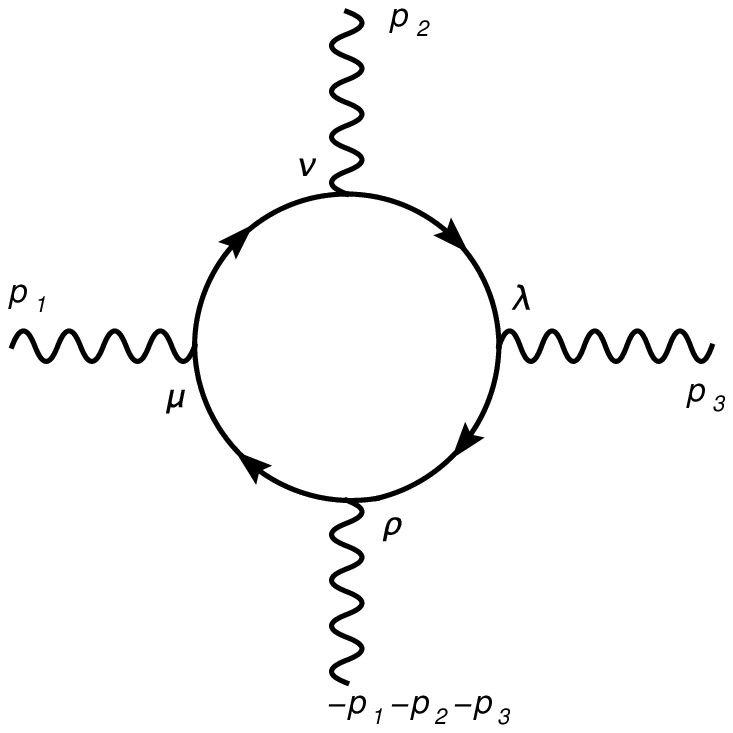}
}\qquad\quad\quad\,\ ,
\end{displaymath}
plus all the equivalent diagrams with permutations of the external legs (one leg must be kept fixed to avoid double counting). A straightforward calculation gives us
\be
\mathcal{L}_{eff}^{(1)}=\frac{2}{45}\frac{\alpha^2}{m^4}(4\mathcal{F}^2 + 7\mathcal{G}^2) \ ,
\ee
where the superscript $(1)$ reminds us that this is a one-loop calculation, and
\be
\mathcal{F}=\frac{1}{2}(\vec{E}^2-\vec{B}^2)=-\frac{1}{4}F_{\mu\nu}F^{\mu\nu}\quad\quad\quad\mathcal{G}=\vec{E}\cdot\vec{B}=-\frac{1}{4}F_{\mu\nu}\widetilde{F}^{\mu\nu} \ ,
\ee
where $F_{\mu\nu}=\partial_{\mu}A_{\nu}-\partial{\nu}A_{\mu}$, and $\widetilde{F}^{\mu\nu}= \frac{1}{2}\epsilon^{\mu\nu\rho\sigma}F_{\rho\sigma}$.

\section{Effective lagrangians for the interaction with a background field}
\label{B}

\begin{figure}[hhh]
\centering
\includegraphics[width=0.29\columnwidth]{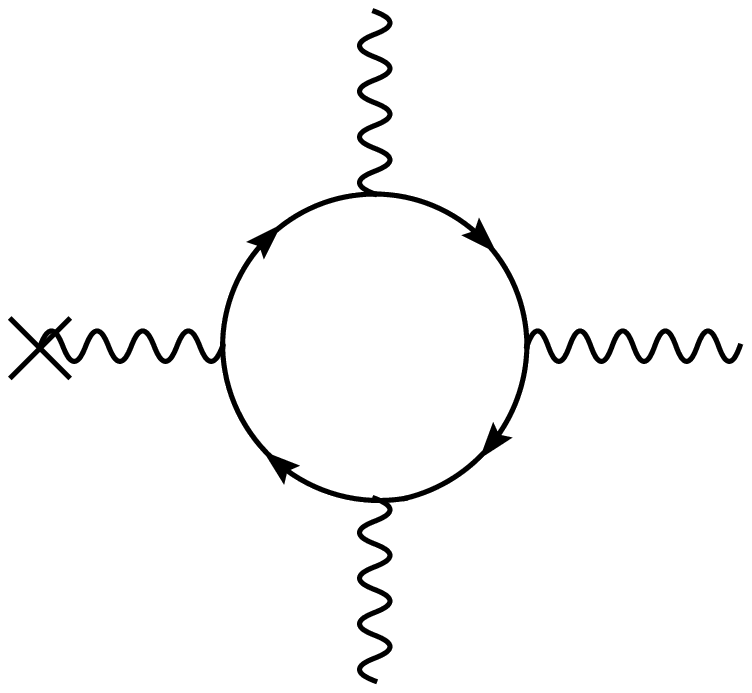}
\includegraphics[width=0.29\columnwidth]{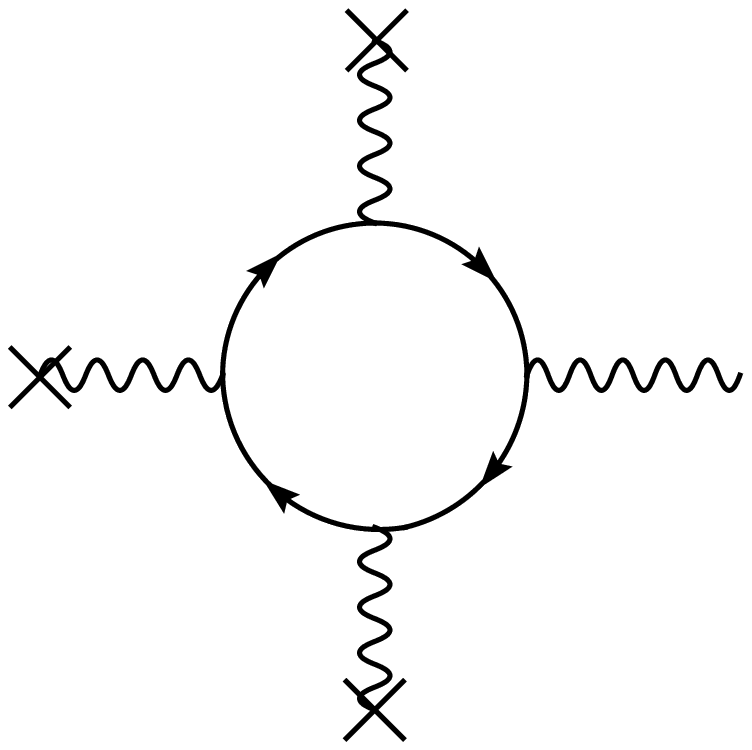}
\caption{\label{ext1-3}Interaction with one and three background-field photons.}
\end{figure}

The Euler-Heisenberg lagrangian describes at lowest order the self-interaction of the electromagnetic field due to the vacuum polarization . Modifying it to include interactions with a background field is straightforward. 
\begin{figure}[hhh]
\centering
\includegraphics[width=0.29\columnwidth]{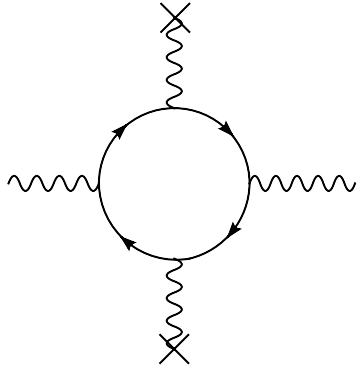}
\caption{\label{ext2}Interaction with two background-field photons.}
\end{figure}
Consider the case in figure \ref{ext2},
corresponding to the leading-order interaction of the electromagnetic field with two photons of the background field. The lagrangian that describes this process has the same form as the Euler-Heisenberg lagrangian, replacing two of the fields by the (classical) background field in all three possible independent combinations. That is,
\bea
\label{L1}
\mathcal{L}_{eff}^{B,2}&=&  \frac{2}{45}\frac{\alpha^2}{m^4}\left\{ \frac{1}{4}\lp F^{\mu\nu}F_{\mu\nu}  F^{\rho\sigma}_BF_{\rho\sigma}^B +  2 F^{\mu\nu}F_{\rho\sigma} F^{\rho\sigma}_BF_{\mu\nu}^B\rp \right.\nn\\
&&\left.+ \frac{7}{64}\lp \epsilon^{\mu\nu\rho\sigma}\epsilon^{\alpha\beta\lambda\eta}F_{\rho\sigma}F_{\mu\nu} F_{\lambda\eta}^BF_{\alpha\beta}^B + 2 \epsilon^{\mu\nu\rho\sigma}\epsilon^{\alpha\beta\lambda\eta}F_{\rho\sigma}F_{\lambda\eta} F_{\mu\nu}^BF_{\alpha\beta}^B\rp\right\}  \ ,
\eea
where the superscript $B$ stands for ``background", and the superscript ``2" stands for the number of background fields.
The other possible cases can be found immediately. For the interaction with just one photon of the background field, the effective lagrangian is
\be
\label{L2}
\mathcal{L}_{eff}^{B,1} =\frac{2}{45}\frac{\alpha^2}{m^4}\left\{  F^{\mu\nu} F_{\mu\nu} F^{\rho\sigma} F_{\rho\sigma}^B + \frac{7}{16} \epsilon^{\mu\nu\rho\sigma}\epsilon^{\alpha\beta\lambda\eta}F_{\rho\sigma}F_{\mu\nu} F_{\lambda\eta} F_{\alpha\beta}^B\right\}\ ,
\ee
and for the interaction with three background photons the lagrangian is
\be
\label{L3}
\mathcal{L}_{eff}^{B,3} =\frac{2}{45}\frac{\alpha^2}{m^4}\left\{  F^{\mu\nu}  F^B_{\mu\nu} F_B^{\rho\sigma} F_{\rho\sigma}^B + \frac{7}{16} \epsilon^{\mu\nu\rho\sigma}\epsilon^{\alpha\beta\lambda\eta}F_{\rho\sigma} F^B_{\mu\nu} F^B_{\lambda\eta} F_{\alpha\beta}^B\right\}\ .
\ee
The diagrams generating these lagrangians are shown in figure \ref{ext1-3}.

\section{Polarization tensor in the presence of an external background field}
\label{P}
Differentiating the effective lagrangian given in eq. (\ref{L1}) with respect to $A^{\mu}$ and $A^{\nu}$ we can obtain immediately the expression of the first correction to the polarization tensor in the presence of a background field as an effective vertex,
\bea
\delta\Pi^{\mu\nu}(q)&=& -\frac{1}{90}\frac{\alpha^2}{m^4}\left\{ 56\left[ q_{\lambda} F^{\lambda\rho}_BF_{\rho\sigma}^B q^{\sigma} g^{\mu\nu} +  F^{\mu\lambda}_BF_{\lambda\rho}^B g^{\rho\nu}q^2 + F^{\mu\lambda}_BF_{\lambda\rho}^B q^{\rho}q^{\nu}+ F^{\nu\lambda}_BF_{\lambda\rho}^B q^{\rho}q^{\mu}\right]\right.\nn\\
&&\left.\qquad\qquad+24 F^{\mu\lambda}_BF^{\nu\rho}_B q_{\lambda} q_{\rho} - 20  F^{\rho\sigma}_BF_{\rho\sigma}^B(q^2 g^{\mu\nu}-q^{\mu}q^{\nu})\right\} \ . \nn
\eea
\be
\ee
For the case of a constant magnetic field $\vec{B}$, this becomes
\be
\delta\Pi^{\mu\nu}(q) = -\frac{\alpha}{\pi}\beta^2\frac{1}{45}\left[2(g^{\mu\nu}q^2-q^{\mu\nu}) - 7(g^{\mu\nu}q^2-q^{\mu}q_{\nu})_{||} + 4(g_{\mu\nu}q^2-q^{\mu\nu})_{\perp}\right] \ ,
\ee
where $\beta\equiv e B/m^2$ and the subscripts $||$ and $\perp$ indicate the projection to the ``parallel" and ``perpendicular" subspaces, namely $(q_0,\vec{q}_{||})$ and $(0,\vec{q}_{\perp})$, where $\vec{q}_{||}$ and $\vec{q}_{\perp}$ are the three-momenta parallel and perpendicular to $\vec{B}$, respectively. This is exactly the expression used in Ref. \cite{Penin}.

\section{Electromagnetic field generated by a charged spherical shell in the presence of a constant magnetic field in the $\hat{z}$ direction}
\label{D}

We can now use the effective lagrangians we found to compute the field generated by a charged spherical shell in the presence of a constant magnetic field $\vec{B}=B\hat{z}$. This problem was studied recently in Ref. \cite{Dominguez}. They performed a classical calculation, solving Maxwell's equations for the lagrangian $\mathcal{L}=\mathcal{L}_M + \mathcal{L}_{eff}^{(1)}$, where $\mathcal{L}_M$ is the Maxwell lagrangian. Their approach involved solving a coupled system of non-linear differential equations. Using our formalism we can split the problem into different parts,  and solve the equations of motion iteratively. Our lagrangian, including the self-interactions of the field as well as the interactions with the background field and the external source, reads
\be
\mathcal{L}_M= -\frac{1}{4}F^{\mu\nu}F_{\mu\nu} + \mathcal{L}_{eff}^{(1)} + \mathcal{L}_{eff}^{B,1} + \mathcal{L}_{eff}^{B,2} + \mathcal{L}_{eff}^{B,3} - j^{\mu}A_{\mu} \ ,
\ee
where
\be
\ee
represents the charged spherical shell of radius $R$ and charge $Q$. We find the following expressions for the electric and magnetic fields: for $r>R$,
\bea
\label{eqext}
\vec{E}(\vec{r})&=& \frac{Q}{4\pi r^2}\hat{r}-\frac{Q}{4\pi}\frac{\alpha^2}{45m^4} \nabla \lp \frac{B^2}{r}(1+7 \text{Cos}(2\theta)- \frac{7}{3}\frac{B^2R^2}{r^3}(3 \text{Cos}(2\theta)+1) - \frac{1}{10\pi^2}\frac{Q^2}{r^5}\rp\nn\\
\vec{B}(\vec{r})&=& \nabla\times\left[ \frac{\alpha^2}{45m^4}\frac{Q^2}{(4\pi)^2}\frac{\vec{B}\times\vec{r}}{r^3}\lp \frac{1}{r}-\frac{4}{3}\frac{1}{R}\rp\right] \ ,
\eea
and for $r<R$,
\be
\vec{E}(\vec{r})=0 \quad\quad\quad\quad \vec{B}(\vec{r})=-\nabla\times\left[\frac{1}{3}\frac{\alpha^2}{45m^4}\frac{Q^2}{(4\pi)^2}\frac{\vec{B}\times\vec{r}}{R^4}\right] \ .
\ee
We find a disagreement with the result of \cite{Dominguez} in the sign of the $1/R$ term in the expression of $\vec{B}$ in eq. (\ref{eqext}).

\section{Conclusions}
\label{conc}

We have modified the Euler-Heisenberg lagrangian to include the possible cases of the interaction between a test field and a background field. From the effective lagrangians we have thus found it is immediate to obtain the first correction to the polarization tensor in the presence of an external field, or the fields induced around a source by the presence of an external electromagnetic field.

\ack

I would like to thank Andrzej Czarnecki and the organizers of the Lake Louise Winter Institute 2009. This work is supported by Science and Engineering Research Canada.

\end{document}